\documentclass[a4paper,11pt]{article}
\usepackage{pos}

\usepackage{graphicx}
\usepackage{color}
\usepackage{physics}
\usepackage{multirow}
\usepackage{arydshln}
\usepackage{hyperref}
\usepackage{url}
\usepackage{lineno}
\usepackage{float}
\usepackage{amsthm}
\usepackage{booktabs}
\usepackage{amsmath}

\newcommand{\figscale}{0.5}
\newcommand{\MSbar}{$\overline{\mathrm{MS}}$ }

\title{The perturbative computation of the gradient flow coupling for the twisted Eguchi--Kawai model with the numerical stochastic perturbation theory}

\author*[a]{Hironori Takei}
\author[a,b]{Ken-Ichi Ishikawa}
\author[a]{Masanori Okawa}

\affiliation[a]{Graduate School of Advanced Science and Engineering, Hiroshima University,\\
  Higashi-Hiroshima, Hiroshima 739-8526, Japan}

\affiliation[b]{Core of Research for the Energetic Universe, Graduate School of Advanced Science and\\
Engineering, Hiroshima University, Higashi-Hiroshima, Hiroshima 739-8526, Japan
}

\emailAdd{t-hironori@hiroshima-u.ac.jp}

\abstract{
The gradient flow method is a renormalization scheme in which the gauge field is flowed by the diffusion equation.
The gradient flow scheme has benefits that the observables composed of flowed gauge fields 
do not require further renormalization and do not depend on the regularization. 
From the independence of the regularization, this scheme allows us to relate 
the lattice regularization and the dimensional regularization such as the \MSbar scheme.
We compute the gradient flow coupling for the twisted Eguchi--Kawai model 
using the numerical stochastic perturbation theory.
In this presentation we show the results of the perturbative coefficients of the gradient flow coupling and its flow time dependence. 
We investigate the beta function from the flow time dependence and 
discuss the lattice artifacts in the large flow time in taking the large-$N$ limit.
}

\FullConference{
The 41st International Symposium on Lattice Field Theory (LATTICE2024)\\
28 July - 3 August 2024\\
Liverpool, UK\\}

\begin{document}
\maketitle

\section{Introduction}
\label{sec:intro}
The gradient flow scheme was initially introduced within the framework of lattice QCD \cite{Narayanan_2006,L_scher_2010,L_scher_2011} 
and has since gained recognition as a renormalization method applicable to both lattice and continuum theories. 
Establishing a correspondence of QCD operators between the lattice scheme and the \MSbar scheme is crucial, 
as theoretical QCD results are often expressed in the \MSbar (or MS) scheme to enable facilitate comparison with experimental observations. 
The gradient flow serves as an intermediate scheme, effectively bridging the gap between the lattice and \MSbar schemes 
since the gradient flow coupling is independent from the regularization.

Despite its advantages, computing the gradient flow coupling expanded in terms of the lattice bare coupling 
presents significant challenges, as it involves solving the gradient flow equation on the lattice theory. 
To overcome this difficulty, we employed numerical stochastic perturbation theory (NSPT) \cite{Parisi_Wu,Renzo_1994,Renzo_1994_2}, 
which provides an efficient approach for performing lattice perturbation theory analyses.

We compute the gradient flow coupling $\lambda_\rho$ expanded in 
the lattice bare coupling $\lambda_0$ in the SU($\infty$) gauge theory 
by using the NSPT for the twisted Eguchi--Kawai (TEK) model \cite{Eguchi_Kawai,TEK_1983,TEK_1983_lat},
\begin{align}
  \lambda_\rho(\mu) = \lambda_0 + r_1(\hat{t})\lambda_0^2 + r_2(\hat{t})\lambda_0^3 + \cdots,
\label{eq:gf_coupling}
\end{align}
where $\mu$ is energy scale and $\hat{t}=t/a^2$ is dimensionless flow time with the lattice spacing $a$.
The renormalization scale $\mu$ and parameter $\rho$ are related as $\mu^2 t=\rho$.
We investigate the flow time dependence of the coefficients $r_1(\hat{t})$ and $r_2(\hat{t})$
and discussed the determination of the one- and two-loop beta functions.
For more detailed discussions and results of this presentation 
can be found in Refs.~\cite{Ishikawa:2024ubo,Takei_d_thesis}.

\section{Gradient flow coupling with NSPT}
\label{sec:gf_nspt}
We generated the perturbative configuration expanded in the lattice bare coupling $g_0$ 
using the NSPT for the TEK model, detailed in Ref.~\cite{Gonz_lez_Arroyo_2019}.
In the gradient flow scheme, the perturbatively expanded link variables 
$V_\mu(\hat{t}) = \sum_{k=0}^\infty g_0^{k} V_\mu^{(k)}(\hat{t})$ are evolved 
using the following hierarchical equation for the perturbative order $k=1,\cdots,\infty$,
\begin{align}
  \frac{d V_\mu^{(k)}(\hat{t})}{d\hat{t}} =& \frac12 \qty( F_\mu[V;\hat{t}] \star V_\mu(\hat{t}) )^{(k)}
  ,\qquad
  V_\mu^{(k)}(\hat{t}=0) = U_\mu^{(k)},
  \label{eq:flow_equation_nspt} 
\end{align}
where
\begin{align}
  F_\mu^{(k)}[V;\hat{t}] =& \left(S_\mu^{(k)} - S_\mu^{(k)\dag} \right) -\frac{1}{N}\mathrm{Tr}\left(S_\mu^{(k)} - S_\mu^{(k)\dag} \right)
  ,\\
  S_\mu^{(k)} =&
  \left(V_\mu \star \sum_{\nu\neq\mu}\left( z_{\mu\nu}V_\nu \star V_\mu^\dag \star V_\nu^\dag + z_{\nu\mu}V_\nu^\dag \star V_\mu^\dag \star V_\nu\right) \right)^{(k)},
\end{align}
and the $\star$-product represents the convolution product of matrix polynomials.

The definition of the gradient flow coupling $\lambda_\rho$ is given as \cite{ramos_2014_tgf}
\begin{align}
  \lambda_{\rho} = \frac{1}{\mathcal{N}(\hat{t})}\expval{\frac{\hat{t}^2 E(\hat{t})}{N}}, 
  \qquad
  E(\hat{t}) =& \frac{1}{2} \sum_{\mu\neq\nu}\Tr\qty( G_{\mu\nu}[V;\hat{t}]G_{\mu\nu}[V;\hat{t}] ),
  \label{eq:def_lgf}
\end{align}
where the field strength tensor $G_{\mu\nu}$ is 
\begin{align}
  G_{\mu\nu}[V;\hat{t}] \equiv& -\frac{i}{8}\qty(C_{\mu\nu}[V;\hat{t}] - C_{\mu\nu}^\dag[V;\hat{t}] ),
  \\
  C_{\mu\nu}[V;\hat{t}] \equiv& z_{\mu\nu}\qty( V_\mu V_\nu V_\mu^\dag V_\nu^\dag + V_\mu^\dag V_\nu^\dag V_\mu V_\nu  + V_\nu^\dag V_\mu V_\nu V_\mu^\dag  + V_\nu V_\mu V_\nu^\dag V_\mu^\dag ).
\end{align}
The perturbative coefficients $r_i(\hat{t})$ of the gradient flow coupling Eq.~\eqref{eq:gf_coupling} can be evaluated by using 
the configurations, flowed through the hierarchical flow equation Eq.~\eqref{eq:flow_equation_nspt}.

\section{Numerical result}
\label{sec:result}
We generated perturbative configurations through the NSPT simulations.
The simulation parameters for the NSPT and the TEK model are summarized in Table~\ref{tab:nspt_parameter}. 
We employed the fixed trajectory length $\tau_L=1.0$ for momentum refreshment,
using the partial refreshment parameter $c_1=\exp(-\gamma\Delta\tau)$ with $\gamma=2.0$, as detailed in Ref.~\cite{Gonz_lez_Arroyo_2019}.
Additionally, the leapfrog integration step size of $\Delta\tau=1/32$ was applied throughout the simulations.
Each sample is separated by $4\tau_L$.
The simulations were conducted using three matrix sizes, $N=289$, $441$, and $529$, for the SU($N$) TEK model. 
The phase parameter was set at $\theta\simeq0.40$ to ensure a smooth approach to the large-$N$ limit~\cite{Gonz_lez_Arroyo_2010}. 
The numerical integration of the gradient flow equation was performed using the L\"uscher scheme~\cite{L_scher_2010},
with the finite step size of $\epsilon=0.01$.
The systematic errors of the numerical integration were negligible as discussed in Ref.~\cite{Ishikawa:2024ubo}.

\begin{table}[h]
  \centering
  \begin{tabular}{cccccccc}
  \toprule
   $\hat{L}$ & $N$ & $k$ & $|\bar{k}|$ & $\theta=|\bar{k}|/L$ & $\tau_L$ & $N_{\mathrm{MD}}$ & Statics
   \\ \midrule
   17 & 289 &  5 &  7 & 0.41 & 1.0 & 32 & 5931 
   \\
   21 & 441 & 13 &  8 & 0.38 & 1.0 & 32 & 3790 
   \\
   23 & 529 &  7 & 10 & 0.43 & 1.0 & 32 & 2600 
   \\ 
   \bottomrule
  \end{tabular}
  \caption{Parameters for TEK model and the NSPT simulations.}
  \label{tab:nspt_parameter}
\end{table}

\subsection{Flow time dependence}
\label{subsec:flow}
We computed the perturbative coefficients $r_1(\hat{t})$ and $r_2(\hat{t})$ for the gradient flow coupling
in Eq.~\eqref{eq:def_lgf}.
In the large flow time region, the finite volume effect of $\order{\hat{t}^2/N^2}$ becomes significant.
Therefore, we determined the beta function using results for $\hat{t}\leq6.3$.
Figure~\ref{fig:r1_largeN} illustrates the flow time dependence for the one-loop coefficient $r_1(\hat{t})$.
The black dashed lines represent the analytic curves~\cite{L_scher_1995,Harlander_2016}, 
while the red crosses with bars correspond to the NSPT results.
To extract the beta function we fitted using the following functions 
in two regions $\hat{t}\in[0.9,6.3]$ and $[2.1,6.3]$,
\begin{align}
  f(\hat{t}) = B_0\qty(\log\qty(\sqrt{2\hat{t}}) + \frac{\gamma_E}{2}) + F_1,
  \quad
  g(\hat{t}) = B_0\qty(\log\qty(\sqrt{2\hat{t}}) + \frac{\gamma_E}{2}) + F_1+\frac{A_0}{\hat{t}},
\label{eq:fit_function_r1}
\end{align}
where $B_0,F_1$ and $A_0$ are fitting parameters.
The fitting results are in Tables~\ref{tab:r1_fit_midle} and \ref{tab:r1_fit_small}.
In the small flow time region, the lattice spacing error $\order{a^2/t}$ becomes large.
The fitting performed with the function $g(\hat{t})$, which incorporates the lattice spacing error
by the term with $A_0$ and includes small flow time $\hat{t}\in[0.9,6.3]$, was found to accurately reproduce 
the results for both the one-loop beta function $\beta_0$ and the constant $f_1$ with good precision.

Figure~\ref{fig:r2_largeN} is the same as Fig.~\ref{fig:r1_largeN}, but the results is subtracted 
the two-loop coefficient $r_2(\hat{t})-r_1(\hat{t})^2$ (crosses) in the large-$N$ limit
and the fit function (solid line) is
\begin{align}
  f(\hat{t}) = B_1\qty(\log\qty(\sqrt{2\hat{t}}) + \frac{\gamma_E}{2}) + F_2,
\label{eq:fit_function_r2r1}
\end{align}
with the fitting parameters $B_1$ and $F_2$.
The results of fitting are summarized in Table~\ref{tab:r2_fit_largeN}.
However, the large errors prevent us from precise determination of the two-loop beta function coefficient.
In order to accurately extract the two-loop beta function coefficient, it is necessary to
further minimize the statistical error in the large-$N$ limit.

To reduce statistical errors, one approach is to increase the number of the samples.
Additionally, incorporating data points at larger $N$ could enhance the stability of 
the flow time fitting by extending the flow time region where the large-$N$ limit can be 
reliably extrapolated with controlled $\order{\hat{t}^2/N^2}$ corrections.
Moreover, leveraging the large-$N$ factorization property could reduce the variance of the coefficients,
thereby minimizing statistical errors at larger $N$.
In the following subsection, we investigate whether the large-$N$ factorization can be verified for the 
flowed operator by examining the variance of the perturbative coefficients.

\renewcommand{\figscale}{0.455}
\begin{figure}[t]  
  \centering
      \includegraphics[clip,scale=\figscale]{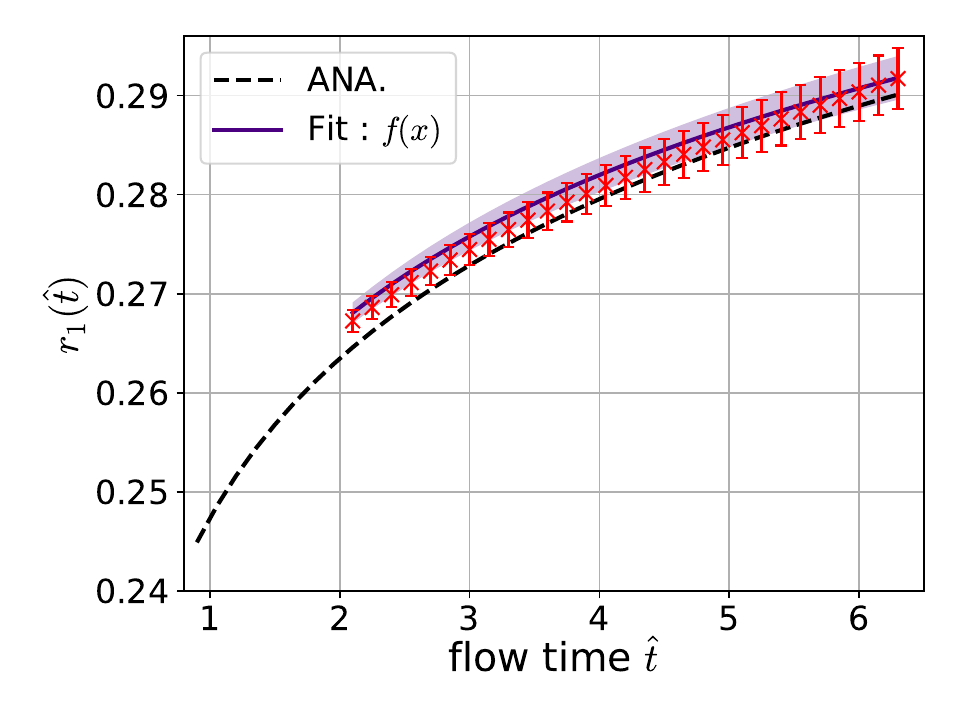}\hfill
      \includegraphics[clip,scale=\figscale]{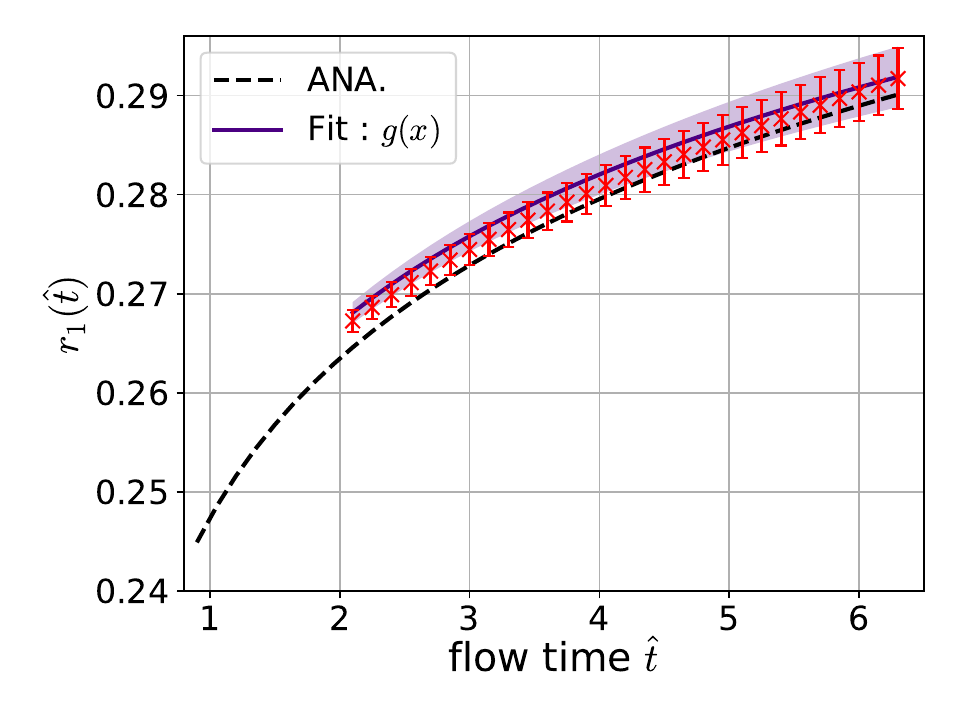}
      \includegraphics[clip,scale=\figscale]{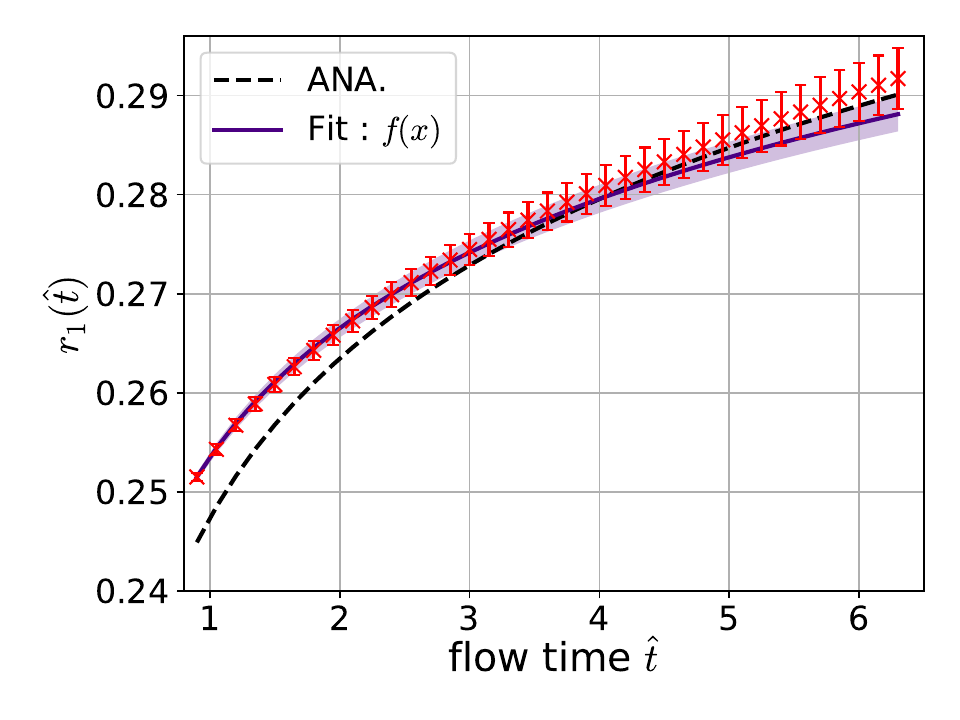}\hfill
      \includegraphics[clip,scale=\figscale]{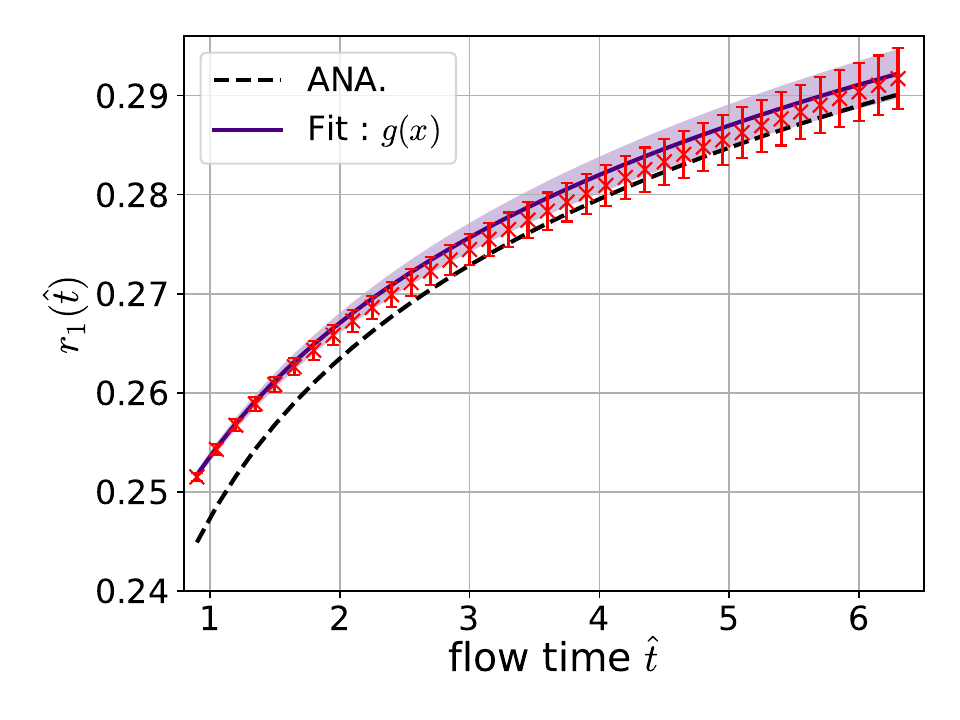}
      \caption{Flow time dependence of the one-loop coefficient $r_1(\hat{t})$ in the large-$N$ limit.
      Top (bottom) panels present the fitting results in the flow time region $\hat{t}\in[2.1,6.3]$ ($\hat{t}\in[0.9,6.3]$).
      Red crosses represent simulation results, while black dashed lines correspond to continuum analytic expressions.
      Purple solid lines represent the fitting results with $f(x)$ (left panels) and $g(x)$ (right panels).
      }
      \label{fig:r1_largeN}
\end{figure}

\begin{table}[h]
  \centering
  \caption{Fitting results of $r_1(\hat{t})$ in $\hat{t}\in[2.1,6.3]$.
  The results are plotted in Fig.~\ref{fig:r1_largeN}.}
  \begin{tabular}{cllll} \toprule
      Fit function & $B_0$ & $F_1$ & $A_0$ & $\chi^2/N_\mathrm{dof}$\\ \midrule
      \multirow{1}{*}{$f(x)$} 
            & 0.04315(222) & 0.22466(139)  & --           & 12.3 \\
      \multirow{1}{*}{$g(x)$} 
            & 0.04349(748) & 0.22419(1009) & 0.00030(634) & 12.1 \\ \midrule
      Analytical value & 0.046439     & 0.217862 & -- & -- \\
  \bottomrule
  \end{tabular}
  \label{tab:r1_fit_midle}
\end{table}

\begin{table}[h]
  \centering
  \caption{Same as Table.~\ref{tab:r1_fit_midle}, but the fit region is $\hat{t}\in[0.9,6.3]$.}
  \begin{tabular}{ccllll} \toprule
      Fit function & $B_0$ & $F_1$ & $A_0$ & $\chi^2/N_\mathrm{dof}$\\
      \midrule
      \multirow{1}{*}{$f(x)$}
       & 0.03762(144) & 0.22959(60) & -- & 3.2 \\
      \multirow{1}{*}{$g(x)$}
       & 0.04725(349) & 0.21778(337) & 0.00577(139) & 4.2 \\ \midrule
      Analytical value & 0.046439 & 0.217862 & -- & -- \\
  \bottomrule
  \end{tabular}
  \label{tab:r1_fit_small}
\end{table}

\renewcommand{\figscale}{0.55}
\begin{figure}[t]  
    \centering
        \includegraphics[clip,scale=\figscale]{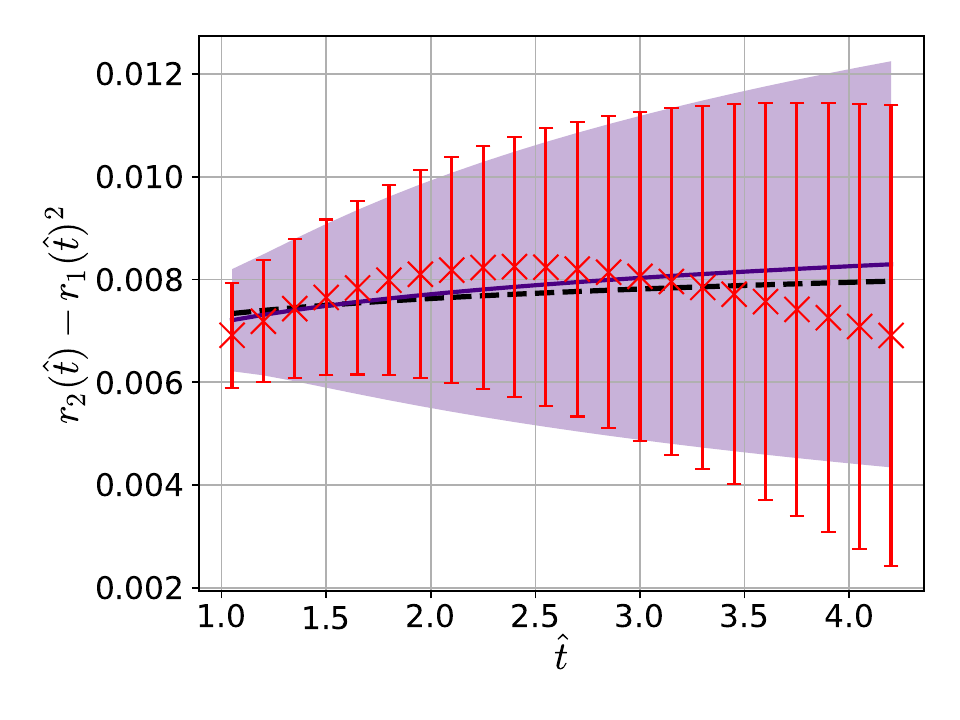}
        \caption{Same as Fig.~\ref{fig:r1_largeN}, but for $r_2(\hat{t})-r_1(\hat{t})^2$.}
        \label{fig:r2_largeN}
\end{figure}

\begin{table}[t]
    \centering
    \caption{Fitting parameters of the coefficient $r_2(\hat{t})-r_1(\hat{t})^2$ in the flow time region $\hat{t}\in[2.1,6.3]$.
    An uncorrelated fit was used in this case.
    }
     \begin{tabular}{clll}\toprule
        Fit function  &    $B_1$       & $F_2$        & $\chi^2/N_\mathrm{dof}$\\ \midrule
            $f(x)$    &   0.00157(491) & 0.00617(293) & 0.032                  \\ \midrule
     Analytical value &   0.00090897   & 0.00673711   & -- \\  
    \bottomrule
    \end{tabular}
    \label{tab:r2_fit_largeN}
\end{table}

\subsection{Large-$N$ factorization}
\label{subsec:fact}
The factorization property is a fundamental aspect of SU($N$) gauge theory in the large-$N$ limit, 
signifying that the variance of an operator vanishes as $N\to\infty$. 
Previous studies \cite{Gonz_lez_Arroyo_2019, Gonz_lez_Arroyo_2022} have demonstrated the large-$N$ factorization 
in the expectation values of perturbative coefficients within matrix models. 
However, when considering flowed operators at finite flow times, the applicability of large-$N$ factorization 
becomes less straightforward and warrants further investigation.
In this subsection, we analyze the variance of these coefficients at finite $N$ at finite flow time
and confirm the existence the large-$N$ factorization for the flowed operator.

Figure~\ref{fig:variance_t6} illustrates the variance of the coefficients 
$r_1(\hat{t})$ and $r_2(\hat{t})$ at the flow time $\hat{t}=6.0$.
In the figures, the red diamonds are the results of the NSPT with finite $N=289$, $441$, and $529$.
The red solid lines correspond to a simple linear fitting using the function $f(N)=A_0+A_1/N^2$.
We observe that the variance at the large-$N$ limit, represented by the red down-triangle, remains finite, 
which at first glance appears to contradict the large-$N$ factorization property. 
However, this does not constitute a true contradiction, as finite-volume corrections of $\order{\hat{t}^4/N^4}$ 
can become significant at $\hat{t}=6.0$ for the smaller values of $N=289$, $441$, and $529$. 
Such finite-volume effects are known to grow with increasing flow time.
This phenomenon is further elucidated in Refs.~\cite{Ishikawa:2024ubo,Takei_d_thesis}, 
which provide a detailed tree-level analysis of the energy density operator.
Consequently, a simple linear extrapolation does not yield zero variance at large-$N$ from our data sets.
To account for the next leading finite volume correction $\order{\hat{t}^4/N^4}$ 
we performed the global fit in $\hat{t}\in[3.0,7.5]$ with the function $f(\hat{t},N)=A_0+A_1\hat{t}^2/N^2+A_2\hat{t}^4/N^4$, 
shown by the black dash-dotted lines in Fig.~\ref{fig:variance_t6}.
The extrapolated variance (black up-triangle) at the large-$N$ limit, $A_0$, was found to be smaller than that obtained from the simple linear fitting.

These results confirm the existence of the large-$N$ factorization at finite flow times.
Therefore, performing NSPT simulations with larger matrix sizes is expected to enable stable fitting with fewer statistics.
An estimate of the required sample size for arbitrary matrix sizes and relative errors is provided in Ref.~\cite{Ishikawa:2024ubo}.

\renewcommand{\figscale}{0.46}
\begin{figure}[h]  
    \centering
    \includegraphics[clip,scale=\figscale]{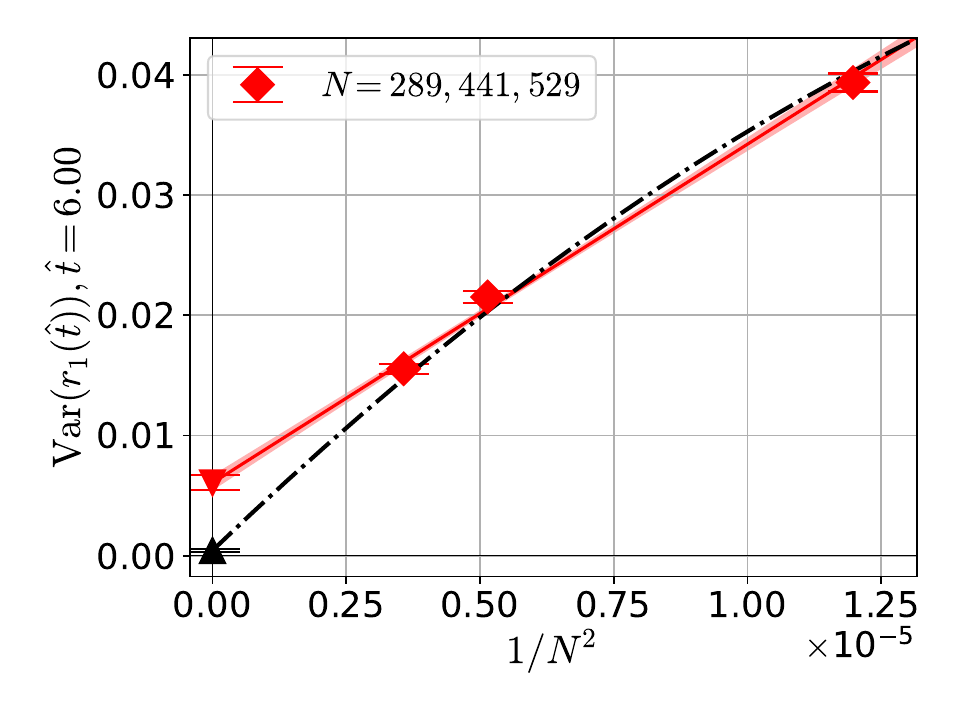}  \hfill
    \includegraphics[clip,scale=\figscale]{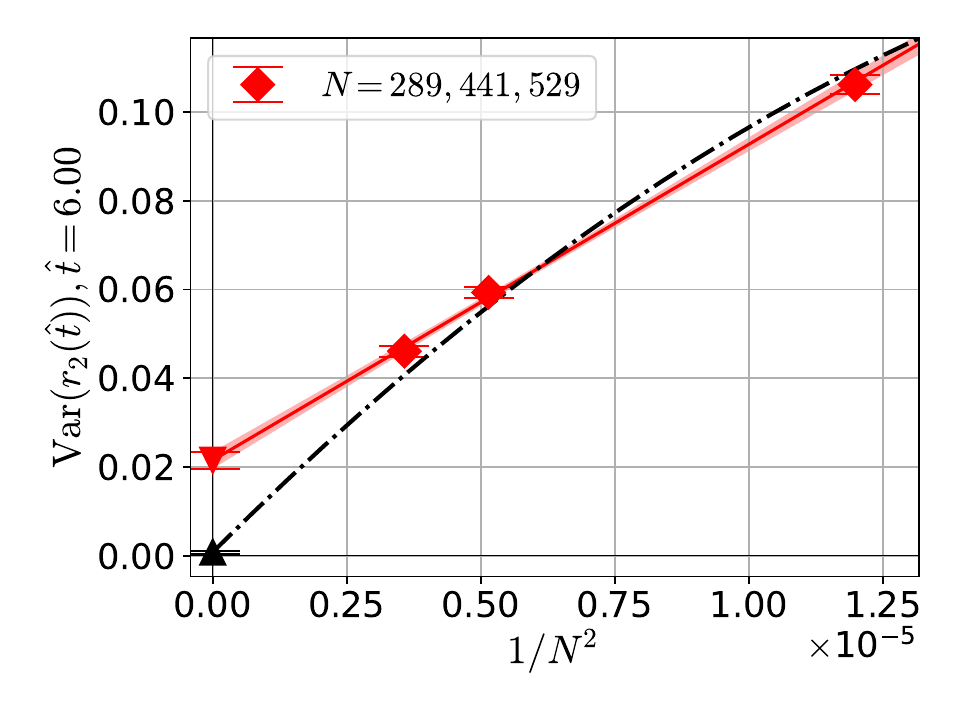}
    \caption{
    Variances of the coefficients $r_1(\hat{t})$ (left panel) and $r_2(\hat{t})$ (right panel) at $\hat{t}=6.0$.
    The red diamonds indicate the numerical results of the variances for the finite $N=289$, $441$, and $529$.
    The red down-triangles denote the results for large-$N$ obtained using simple linear fitting (red solid line).
    }
    \label{fig:variance_t6}
\end{figure}

\section{Summary}
\label{sec:summary}
In this presentation, we evaluated the perturbative coefficients of the gradient flow coupling constant 
expanded in the lattice coupling constant for the large-$N$ gauge theory using NSPT for the TEK model. 
By examining the scale dependence of the coefficients, systematic errors such as 
lattice spacing effects and finite volume effects were identified, 
and the precision with which the beta function can be determined was discussed. 
With the current data sets, the one-loop beta function was determined with an accuracy of less than 10\%.
However, more statistics are required to precisely determine the beta function at two loops or beyond. 
In the latter part, the existence of large-$N$ factorization at finite flow times was confirmed. 
Therefore, calculating the perturbative coefficients with larger matrix sizes is expected to 
allow for more precise determination of the two and higher order beta functions 
with fewer statistical samples.

\section*{Acknowledgements}
H.T. is supported by JST, the establishment of university fellowships toward the 
creation of science technology innovation, Grant Number JPMJFS2129.
K.-I.I. is supported in part by MEXT as “Feasibility studies for the next-generation computing
infrastructure”. M.O. is supported by JSPS KAKENHI Grant Number 21K03576.
The computations were performed using the following computer resources; 
Cygnus at the Center for Computational Sciences, University of Tsukuba,
ITO subsystem-B offered under the category of general projects 
by the Research Institute for Information Technology, Kyushu University,  
and SQUID at the Cybermedia Center, Osaka University under the support of the RCNP joint use program.

\bibliographystyle{ws-ijmpa}
\bibliography{biblio}

\end{document}